\documentclass[10pt,letterpaper]{article}
\pdfoutput=1
\usepackage[top=0.85in,left=2.75in,footskip=0.75in]{geometry}
\pdfoutput=1

\usepackage{amsmath,amssymb}

\usepackage{changepage}

\usepackage[utf8x]{inputenc}

\usepackage{textcomp,marvosym}

\usepackage{cite}

\usepackage{nameref,hyperref}

\usepackage[right]{lineno}

\usepackage{microtype}
\DisableLigatures[f]{encoding = *, family = * }

\usepackage[table]{xcolor}

\usepackage{array}

\newcolumntype{+}{!{\vrule width 2pt}}

\newlength\savedwidth

\raggedright
\usepackage[aboveskip=1pt,labelfont=bf,labelsep=period,justification=raggedright,singlelinecheck=off]{caption}

\makeatletter
\renewcommand{\@biblabel}[1]{\quad#1.}
\makeatother

\usepackage{lastpage,fancyhdr,graphicx}
\usepackage{epstopdf}
\fancyheadoffset[L]{2.25in}
\fancyfootoffset[L]{2.25in}
\usepackage{amsmath}
\usepackage{dcolumn}
\usepackage{siunitx}
\usepackage{wasysym}
\usepackage{multirow}
\usepackage[T1]{fontenc}

\begin{document}
\vspace*{0.2in}

\begin{flushleft}
{\Large
\textbf\newline{Measurement of the dependence of ultra diluted gas transmittance on the size of the detector}
}
\newline
\\
Jakub M. Ratajczak
\\
\bigskip
Centre of New Technologies, University of Warsaw, Poland
\\
\bigskip

j.ratajczak@cent.uw.edu.pl

\end{flushleft}

\section*{Abstract}
We show that measured optical transmittance of an ultra thin gas depends on the detector size. To this end we conducted an experiment that compares transmittances measured in parallel with a pair of detectors with different diameters ranging from \SI{2}{\micro\metre} to \SI{200}{\micro\metre}. A Tunable Diode Laser Absorption Spectroscopy type system was used. Transmittance of $\sim$\SI{e-2}{\milli\bar} water vapor on NIR absorption line $\lambda$=\SI{1368.60}{\nano\metre} was measured using a \SI{60}{\metre} long multi-pass cell placed inside the \SI{300}{\l} vacuum chamber. The result of the experiment shows higher transmittances when the measurement is performed using smaller detectors. The difference reaches as much as 1.23\,$\pm$0.1\,\%, which is greater than 0 with >5$\sigma$ statistical significance. Qualitatively it is in agreement with the recently developed model of thin gas optical transmittance taking into account the quantum mechanical effects of spreading of the wave functions of individual gas particles.

\section*{Introduction}
The Beer-Lambert exponential transmission law \cite{Bouguer1729} \cite{A.D.McNaught1997} describing attenuation of monochromatic light by the homogeneous, not very dense medium is well known for almost three centuries. Despite developing newer, more advanced transmittance models, today it still applies to quantitative spectroscopy \cite{Bernath2016} and rarefied gases, among others. All this models relies on an assumption of attenuating particles locality. However, an increasing number of experiments \cite{Handsteiner2017} \cite{Rauch2018} convince us that the underlying theory of Quantum mechanics is not a local realistic theory \cite{Einstein1935} \cite{Musser2016}. There is one more assumption in most of "classic" transmittance models: a light detector is a macroscopic apparatus. Quantum mechanics is considered to be one of the most fundamental theories so it is necessary to check whether these two assumptions limit scope of applicability of classic models. 

Quantum spreading is an effect that involves spatial smearing of the $\Psi$ wave function over time. It leads to the spreading of the $|\Psi|^2$ density of the probability of any reaction (quantum measurement) of a physical object described by such a function. It is derived from solving the Schrödinger equation for a free particle \cite{Shankar2011}. We applied this solution to each gas particle independently during its free time between successive collisions. It is a kind of "smeared gas". It leads, together with the assumption of non-locality, to a new model \cite{Ratajczak2019b} of electromagnetic transmittance of thin gases. One of the predictions of this model is that the measured optical transmittance depends, among others, on the size of the detector used and the duration of the particles mean free time. The classical, "local" approach to transmittance, the Beer-Lambert law included, does not predict any of such dependencies.

Currently, one of the most popular methods of quantitative testing the transmittance of thin gases is Tunable Diode Laser Absorption Spectroscopy (TDLAS) \cite{Wang2018}. Our setup is a slightly modified version of such a system. The transmittance is measured simultaneously using a pair of detectors with different effective active cross-sections. One laser beam passes through the gas and eventually it’s split towards two detectors with a beam-splitter. There is a big chamber used to provide conditions for a significant spreading of the gas particles wave functions to occur. The model applies to any type of gas. We chose to test the water vapor.

The objective of the experiment is to examine the qualitative prediction on whether the transmittance measured with a smaller detector is greater than the transmittance measured with a larger detector. Results of the experiment confirm this with >5$\sigma$ statistical significance.

The experimental setup is described in the next section. The way data is acquired and processed is covered in the following sections. Results are discussed in the 5th section. Conclusions are presented at the end of the paper. The measurement uncertainty is analyzed in the appendix next to the online data guide.

\section*{Experiment}

\subsection*{Setup}
\begin{figure*}[ht]
\includegraphics[width=\textwidth]{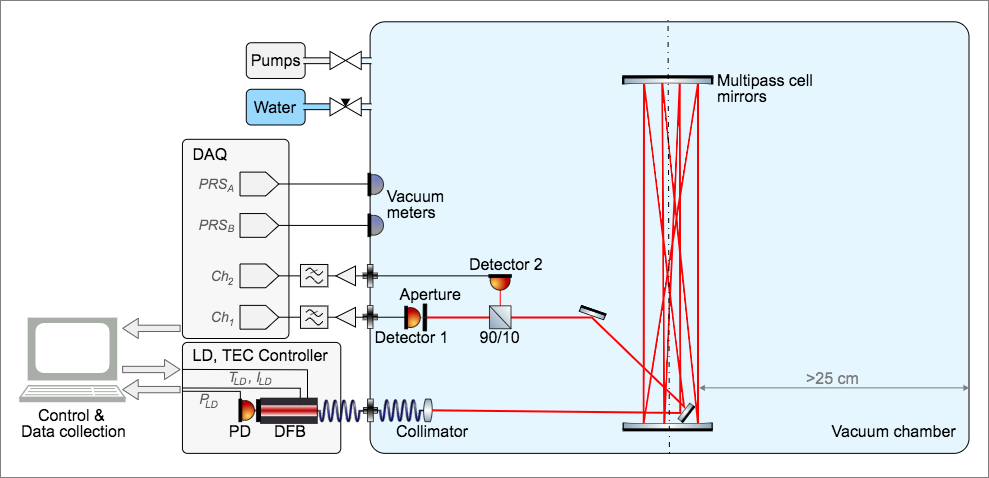}
\caption{The schematic diagram of the experimental setup. All optics and detectors are placed inside the vacuum chamber. The electronic equipment and the laser are kept outside.} \label{fig:schematic_diagram}
\end{figure*}

\begin{figure}[!]
\includegraphics[width=\columnwidth]{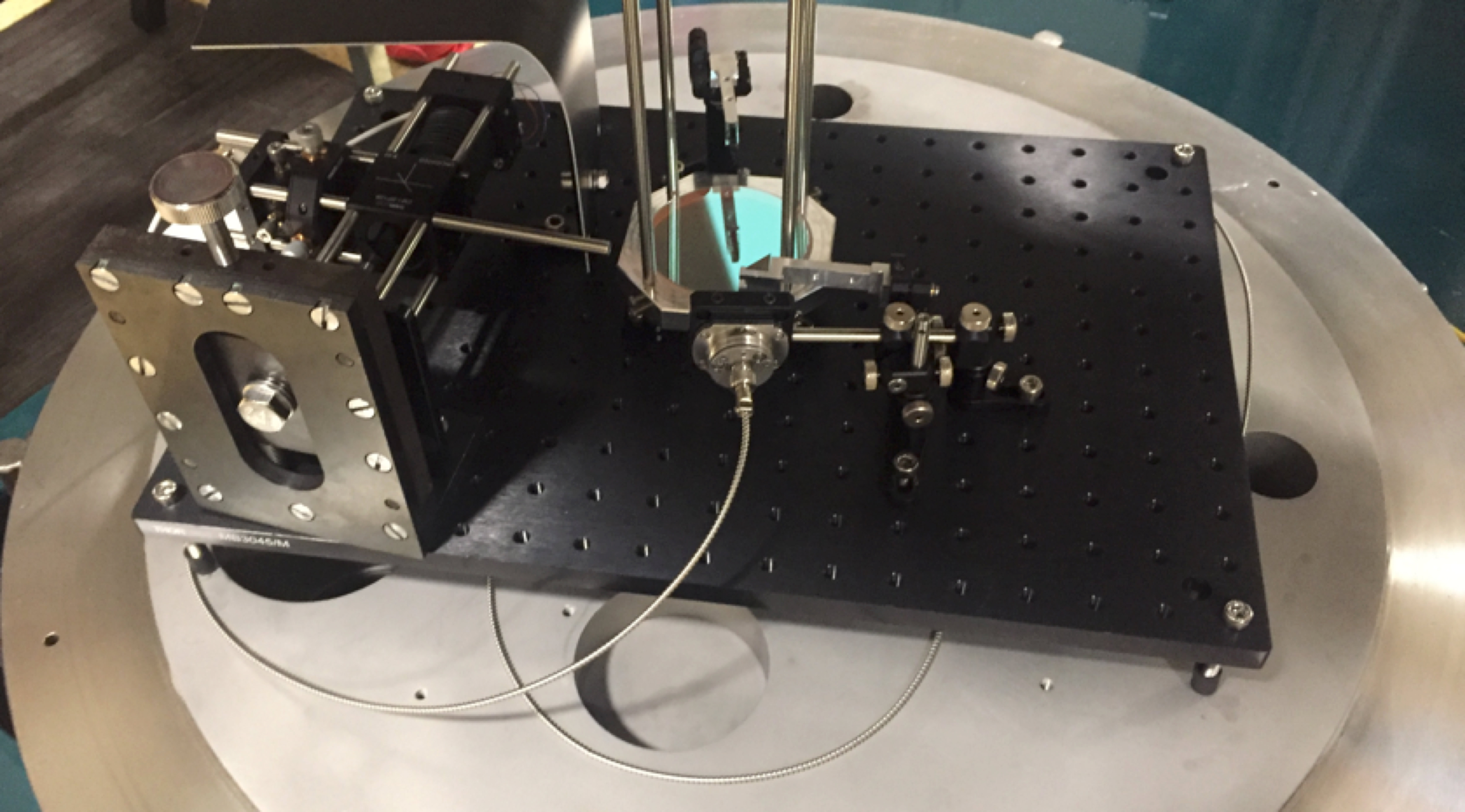}
\caption{A picture of optics in the open chamber.} \label{fig:breadboard_picture} 
\end{figure}

The schematic diagram is shown in FIG.~(\ref{fig:schematic_diagram}). The tested gas, along with the optical setup and the detectors, is placed inside a bell-type vertical cylindrical vacuum chamber ($\diameter$=\SI{58}{\cm}, h=\SI{110}{\cm}) in room temperature. The working pressure is approx. $10^{-3}-$\SI{e-2}{\milli\bar}. A DFB SM-pigtailed laser diode (LD-PD Inc., PL-DFB-1368-A-A81-SA) operating at $\sim$\SI{1368.60}{\nano\metre} with \SI{2}{\MHz} line width is used to scan 101313–000212  \cite{Toth1994} water absorption line. The DFB CW output power is about \SI{1}{\mW} at \SI{27}{\celsius}. The laser diode is equipped with an internal InGaAs photodiode. The DFB diode is controlled by an integrated current and temperature controller (Thorlabs, CLD1015) placed outside the chamber. The temperature setting of the diode is set to \SI{27}{\celsius}. The laser beam is fed into the chamber using single-mode optical fibers (SMF-28E and SMG652.D) using vacuum feedthrough (SQS, KF40 SM). The beam is collimated with adjustable ECO-550 lenses (Thorlabs, PAF2-2C). The vertical multi-pass setup is made up of two 3" dielectric concave mirrors f=\SI{500}{\mm} (Thorlabs, CM750-500-E04-SP) held vertically by metal rods about \SI{75}{\cm} apart from each other. Two flat silver mirrors (Thorlabs, MRA03-P01, MRA10-P01) are used to direct laser beam from the collimator to/from the multipass cell and towards a 90/10 pellicle beamsplitter (Thorlabs, CM1-BP108). The multipass cell axis is in line with the chamber vertical axis to assure equally long distance from the chamber walls, approx. \SI{25}{\cm}. All optical components are installed on a small breadboard standing on the chamber floor, see FIG.~(\ref{fig:breadboard_picture}).

Transmittance is measured in parallel by a pair of $\diameter$\SI{200}{\micro\metre} InGaAs photodiodes (Hamamatsu G11193-02R). A circular aperture stop to reduce detector's effective active area is inserted in front of the detector ($Ch_1$) placed in line with the beamsplitter incoming beam, see Table~(\ref{tab:apertures}). The aperture is installed $\sim$\SI{1}{\mm} away from the photosensitive surface. The second detector ($Ch_2$), without an aperture, gives a reference transmittance reading. Both detectors and the aperture stop are installed in their own X-Y translation units. The $\sim$\SI{90}{\percent} of the beam is directed towards this detector. 

\begin{table*}
\begin{adjustwidth}{-2.25in}{0in} 
\caption{\label{tab:apertures}$Ch_1$ aperture stops parameters.}
\begin{tabular}{rccccccccc}
\hline
Size & \SI{2}{\micro\metre} & \SI{5}{\micro\metre} & \SI{15}{\micro\metre} & \SI{25}{\micro\metre} & \SI{40}{\micro\metre} & \SI{50}{\micro\metre} & \SI{100}{\micro\metre} & \SI{150}{\micro\metre} \\
Model & P2H & P5D & P15D & P25D & P40D & P50D & P100D & P150D \\
Diameter tolerance & $\pm$\SI{0.25}{\micro\metre} & $\pm$\SI{1}{\micro\metre} & $\pm$\SI{1.5}{\micro\metre} & $\pm$\SI{2}{\micro\metre} & $\pm$\SI{3}{\micro\metre} & $\pm$\SI{3}{\micro\metre} & $\pm$\SI{4}{\micro\metre} & $\pm$\SI{6}{\micro\metre} \\
Circularity & $\ge$\SI{85}{\percent} & $\ge$\SI{90}{\percent} & $\ge$\SI{90}{\percent} & $\ge$\SI{95}{\percent} & $\ge$\SI{95}{\percent} & $\ge$\SI{95}{\percent} & $\ge$\SI{95}{\percent} & $\ge$\SI{95}{\percent} \\
\hline
Gain - resistor (\si{\ohm})  & \SI{50e6} & \SI{10e6} & \SI{51e3} & \SI{51e3} & \SI{51e3} & \SI{51e3} & \SI{2e3} & \SI{2e3}{} \\
\hline
\end{tabular}
\end{adjustwidth}
\end{table*}

Both photodiodes operate in a photovoltaic mode \cite{Ready1997} with no bias voltage. The photodiode current is amplified \cite{Orozco} by a custom-built current-to-voltage amplifier using the low current op-amp (TI, LMC662), followed by a low-pass \SI{10}{\Hz} RC filter. The amplifier gain is controlled by a feedback loop resistor. The $Ch_1$ resistance is adjusted for each aperture individually to keep output voltage within a range of a few dozen mV, see Table~(\ref{tab:apertures}). The $Ch_2$ resistance is constant and equal to \SI{2}{\kilo\ohm}. Gains are chosen to obtain both characteristics curves as linear as possible within the range used for actual measurement. 

Vacuum is evacuated with a scroll pump and a turbo molecular pump. Two independent vacuum meters are used (Leybold, Ceravac CTR 100N, Ionivac ITR90).

All voltage data is acquired by the DAQ unit (LabJack, U6). The digital data, from the DAQ unit and the laser driver, is retrieved via a USB by a PC computer running LabView. A custom built LabView GUI is used to automate the measurement procedure. The LabView text files data is extracted periodically into the SQL database from where it is retrieved for statistical analyses performed by the Mathematica 12 package.

In this paper $Ch_1$ and $Ch_2$ may denote either a respective photodiode or one of measurement channels or a voltage read from one of channels.

\subsection*{Gas and spectral line selection}
We selected water vapor due to its large cross-section in the near infrared. The HITRAN \cite{Gordon2017} and MARVEL \cite{Furtenbacher2007} data was referenced. The large cross-section allows for testing transmittance at low pressures with shorter path of light. Low pressure and a narrow spectral range of the laser allow for isolating a single absorption line with sufficient resolution.

The expected mean free path of water particles for the pressure of $\sim$\SI{e-3}{\milli\bar} is longer than \SI{37}{\cm}. With approximate mean speed in the room temperature of \SI{600}{\meter\per\second} it gives approx. \SI{0.6}{\milli\second} particle mean free time. Plugging this time to the Schrödinger equation free particle solution we calculate the particle wave packet standard deviation to be $\sim$\SI{14.6}{\micro\metre}. This is a conservative calculation using ideal gas laws. 

The chamber size is another restriction on the particle mean free path. There should be enough distance between walls and the measuring laser beam. This is why a chamber diameter of \SI{58}{\cm} was chosen.

\subsection*{Stages of the experiment}

The experiment is carried out in successive, several days long runs, see Table~(\ref{tab:runs}). There is a different aperture stop installed during each run. A run consists of many consecutive either $\sim$\SI{30}{\minute} or $\sim$\SI{10}{\minute} long cycles.

\setlength{\tabcolsep}{2pt}

\begin{table}[ht!]
\begin{adjustwidth}{-2.25in}{0in} 
\caption{\label{tab:runs}Basic statistics of the experiment runs.}
\begin{tabular}{rcccccccccccc}
\hline
Run name & R-3 & R-5 & R-6 & R-7 & R-8 & R-9 & R-10 & R-11 & R-12 & R-13 & R-14 & R-15 \\
Aperture installed & \SI{2}{\micro\metre} & \SI{2}{\micro\metre} & \SI{25}{\micro\metre} & \SI{5}{\micro\metre} & none & \SI{150}{\micro\metre} & none & \SI{15}{\micro\metre} & \SI{40}{\micro\metre} & \SI{25}{\micro\metre} & \SI{50}{\micro\metre} & \SI{100}{\micro\metre}  \\
No. of cycles & 131 & 218 & 71 & 81 & 31 & 268 & 709 & 558 & 645 & 1177 & 1014 & 1012 \\
No. of measurements & 384\,289 & 577\,676 & 183\,212 & 214\,318 & 80\,220 & 695\,544 & 1\,185\,491 & 262\,717 & 202\,953 & 555\,846 & 478\,307 & 479\,172 \\
\hline
\end{tabular}
\end{adjustwidth}
\end{table}

Replacement of an aperture requires opening of the vacuum chamber. After the aperture installation, the $Ch_1$ gain is adjusted by soldering in the appropriate feedback loop resistor. In case of two runs (R-8 and R-10), the aperture stop was removed in order to perform a comparative measurement for two detectors equal (according to manufacturer) in size.

After the chamber has been closed, vacuum is build to $\sim$\SI{e-5}{\milli\bar}, which takes roughly 1-2 days. In some cases, the decision was taken to extend this period to 3-5 days. Water is sucked in from an external tank due to the pressure difference. Dosing is controlled using a needle valve adjusted by a manual micrometer screw. Water vapor is produced as a result of spontaneous boiling of water at low pressure.

Water vapor tends to settle on the inside walls of the chamber and setup components \cite{Berman1996}. This can be seen in the spontaneous pressure drop, starting immediately after the feed valve has been closed. To obtain the working pressure in the order of \SI{e-3}-\SI{e-2}{\milli\bar}, the valve is kept open for a longer period of time (e.g., an hour), while maintaining higher pressure (e.g., \SI{2e-2}{\milli\bar}), see pressure plots in FIG.~(\ref{fig:results}).

The pressure measurement below \SI{e-3}{\milli\bar}, made using the Ionivac vacuum gauge ($PRS_B$), is burdened with quite a large error, in the order of $\pm$\SI{15}{\percent}. However, this range is used solely to build vacuum prior to feeding the water. The total working pressure, i.e. above \SI{e-3}{\milli\bar}, is measured, with the accuracy of $\pm$\SI{0.5}{\percent}, using the Ceravac vacuum gauge ($PRS_A$). The partial pressure of the water ($PRS_{H_2O}$) may be determined by converting the classical transmittance measured with $Ch_2$, see Eq.~(\ref{partialprs}). The partial pressure can be determined in our setup within the range of approx. \SI{1e-3}-\SI{3e-2}{\milli\bar}. 

To purify the chamber atmosphere vacuum may be build and water may be fed many times during a single run without opening the chamber. These operations are visible as discontinuities on pressure plots in FIG.~(\ref{fig:results}).

\section*{Data acquisition}

Within a single run the measurement is made in cycles composed of following steps: 

\begin{enumerate}
\item measuring detectors characteristic curves,
\item tuning in the wavelength of the laser light to the chosen absorption line,
\item measuring both transmittances,
\item measuring detectors characteristic curves again (with the laser current step reversed). 
\end{enumerate}

\begin{figure*}[ht!]
\centerline{\includegraphics[width=\textwidth]{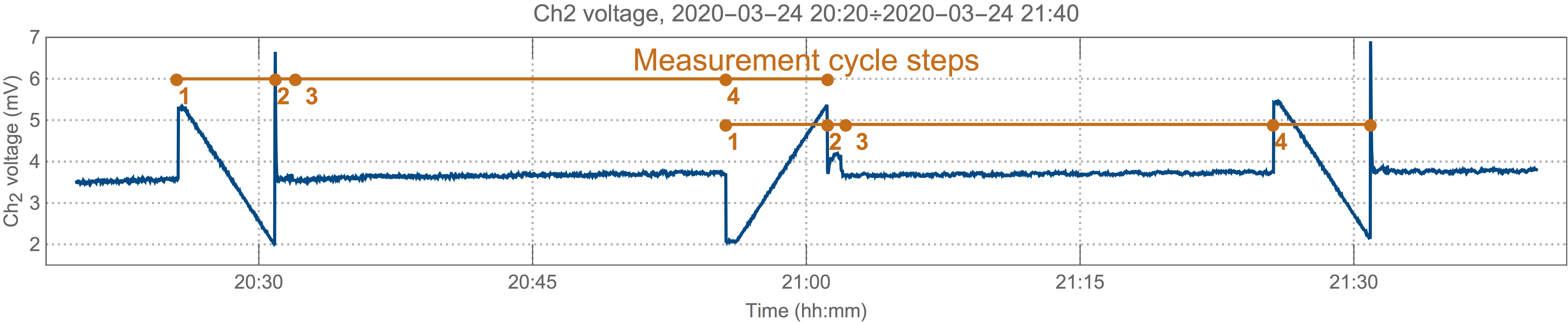}}
\caption{Sample readings of the $Ch_2$ detector during two successive \SI{30}{\minute} cycles along with both cycles steps numbers are presented. Reversing laser current step is visible at the 1st step of the second cycle. } \label{fig:steps}
\end{figure*}

The 4th step is the 1st step of the next cycle. FIG.~(\ref{fig:steps}) presents sample readings of the $Ch_2$ detector during two successive cycles. Cycles last either \SI{30}{\minute} or (in the 2nd part of the experiment) \SI{10}{\minute}. All cyclic measurements are fully automated with a LabView app. Each cycle has its own, time-based identifier assigned e.g., 20200314-0531. The identifier is referred during data post processing and data presentation.

\subsubsection*{Detector characteristics acquisition (step 1)}

The setup is susceptible to various types of drifts: laser tuning, changes in light polarization in optical fibers (affecting the detector readings), pressure and mechanical changes. They cause, among others, a change of the detector characteristics. Due to the adopted method of determining the transmittance quotient, acquiring the appropriate voltage vs. light power characteristic curves is of key importance: $P_1(Ch_1)$ and $P_2(Ch_2)$. Although they are called for convenience “detectors characteristics”, they are not just photodiodes characteristics. They cover all light \& signal processing steps in turn: the fibers \& mirrors light attenuation, the aperture stop ($Ch_2$ only), the photodiodes light-to-current conversion, the current-to-voltage conversion \& amplification and A/D conversion. 

The characteristics are determined by changing the DFB diode current $I_{LD}$. Based on the fact that the DFB diode output power is proportional to the current set, the detector response is determined as a function of the laser power. The range of current changes is automatically selected based on the last reading (from the previous cycle) of the transmittance in the absorption line. It is assumed that the transmittance in the current cycle will be similar to the previous one, i.e. that the partial pressure of the water vapor does not change significantly.

The span of the $I_{LD}$ current changes is arbitrarily set as either ${\Delta}I_{LD}=\pm\SI{12}{\mA}$ for \SI{30}{\min} long cycles or ${\Delta}I_{LD}=\pm\SI{9}{\mA}$ for the shorter ones. DFB controller’s minimum current step is \SI{0.05}{\mA}. It is known that a change of the DFB diode current also causes a change of the wavelength. In case of the diode used in the experiment ${\Delta}{\lambda}/{\Delta}I_{LD}\approx\SI{78}{\pico\metre\per\mA}$, which for \SI{24}{\mA} leads to a change of the wavelength by $\sim$\SI{1.8}{\nm}. We chosen span and range to avoid overlapping with strong absorption lines while characteristics acquisition. 

Such a change in wavelength may also cause a slight change in detector sensitivity (up to \SI{1}{\percent} according to the manufacturer). It is assumed, however, that due to the use of two detectors from the same series, the relative change in sensitivity between the detectors is the same. The same offset cancels possible measurement error because it appears both in the numerator and the denominator of the formula used to determine the transmittance quotient, see Eq.~(\ref{TQ}). 

It is required that the actual laser temperature is within the $\pm$\SI{0.003}{\percent} range from the set point of \SI{27}{\celsius} for each measurement. The scan starts \SI{250}{\ms} after reaching required temperature range. Each recorded measurement within a scan is the arithmetic mean of 30 consecutive single readings ($Ch_1$, $Ch_2$, $P_{LD}$) made every \SI{11}{\ms}. Despite allowing a fairly narrow range of stabilized temperature, the current change direction (up/down) is switched in each cycle. It's either +\SI{0.05}{\mA} or \SI{-0.05}{\mA}. A constant increase (or a decrease, accordingly) of the laser current $I_{LD}$ makes the laser diode working temperature always a little higher (or lower, accordingly) on average. This is because the temperature compensation circuit (Peltier TEC, PID) keeps up with the current changes quite slowly. To compensate this effect the measurements obtained both at the beginning and at the end of the cycle (steps 1 \& 4) are used together to calculate the characteristic curve for the given cycle. This way up to $2{\times}2{\times}12/0.05=960$ (or later $2{\times}2{\times}9/0.05=720$) points are available for approximate characteristics models.

The scan range may be automatically limited from the bottom. For higher pressures (in the order of \SI{e-2}{\milli\bar}) the absorption line is so deep that only a few percent of the light reaches the detectors. In such case the characteristic is obtained close to the lower limit of the laser action and it is impossible to measure some of the distant points in the ${\Delta}I_{LD}$ negative range. The measurement of the characteristics is then limited so that the $I_{LD}$ current fed is large enough (min. \SI{3}{\mA}) to excite the laser action. 

The scan range is limited from the top as well. Laser current  $I_{LD}$ can't exceed \SI{52}{\mA}, so that, by accident, no measurement of the water absorption line is made. 

The characteristic acquisition takes approx. 6 minutes. Most of this time is spent waiting for temperature stabilization of the laser.

\subsubsection*{Tuning in the wavelength (step 2)}

After the characteristics has been measured, a fixed value of the laser current $I_{LD}$ is set to get the wavelength as close as possible to the absorption line maximum. Because of drifts of the diode and the controller it's necessary to adjust the laser current $I_{LD}$. We check, for 3 distinct settings, which transmittance in the $Ch_2$ detector is the lowest: one that is identical as in the previous cycle or one that is different by either plus or minus \SI{0.05}{\mA}. Checking starts 45 seconds after the measurements of the characteristics had been completed, providing time to properly relax temperature of the laser to the range of $\pm$\SI{0.003}{\percent}. This way, actual wavelength is kept as close as possible to the absorption maximum during the entire experiment.

\subsubsection*{The transmittances measurement (step 3)}

Having set the fixed laser current and temperature setpoints near the absorption line maximum, the transmittances measurement is carried out in the continuous mode. We don't use any wavelength modulation techniques. Single measurements are recorded twice a second. A single measurement includes following parameters read synchronously: both detectors voltages ($Ch_1$, $Ch_2$), laser power read from the laser photodiode ($P_{LD}$), actual laser temperature and current ($T_{LD}$, $I_{LD}$) as well as both pressure readings ($PRS_A$, $PRS_B$). This measurement step ends half an hour (or \SI{10}{\min}) from the start of the cycle.

\subsubsection*{The 2nd characteristics acquisition (step 4)}

The last step in the cycle is to acquire measurements for the characteristics again. This time, stepping the laser current in the reverse direction. It is also the first step of the next cycle.

\section*{Data processing}

The difference of the water vapor transmittances measured by both detectors is calculated as the transmittance quotient ($TQ$) according to the following procedure.

\subsection*{Determination of the linear model of the detector characteristics}

The characteristics $P_1(Ch_1)$ and $P_2(Ch_2)$ of both detectors are determined for each measurement cycle. They are approximated with a linear regression near the range of the detector voltage values observed during the given measurement cycle:

\begin{equation}
\begin{aligned}
 P_1(Ch_1)=A_1 Ch_1+B_1~, \\  P_2(Ch_2)=A_2 Ch_2+B_2~,
\end{aligned}
\end{equation}

where $A_{1,2}$ and $B_{1,2}$ are linear model coefficients.

\begin{figure*}[ht!]
\includegraphics[width=\textwidth]{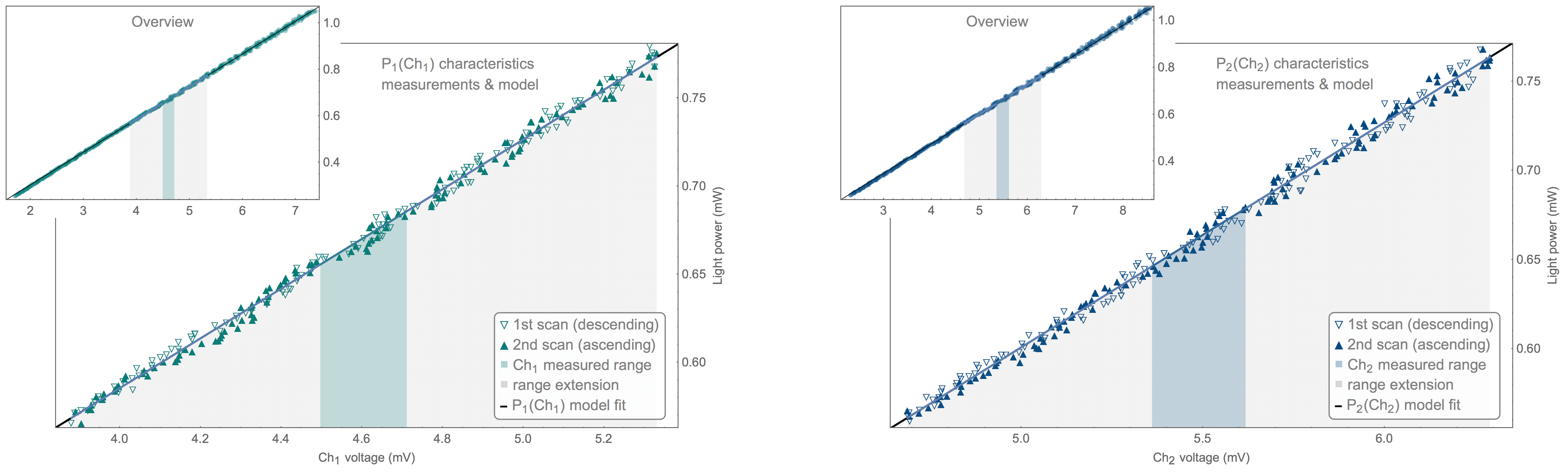}
\caption{Example of linear approximation of both characteristics along with source measurements. The colored vertical stripe indicates the range of the detected voltage variability in the cycle. The grey vertical stripes indicate an additional range (up to \SI{25}{\percent}) of the points used to determine the linear regression.} \label{fig:linear_approximation}

A set of points used for actual fitting is limited to the range of voltages measured by the detector within the cycle. It gives linear approximation with a smaller error within the interesting range. Usually these are very narrow ranges of detector operation (e.g. tenths of \si{mV}), resulting from noise and a small pressure difference during the cycle. Very often they are so narrow that there are too few measurements available for accurate model fitting. The range is then extended up to \SI{25}{\percent} of the voltage range of the characteristic being measured. This way the model is fitted using approx. 150-200 points only. Thanks to the surplus of points, the linear regression can be assessed visually in a broader context, see overview in FIG.~(\ref{fig:linear_approximation}). Such plots are available online for all measurement cycles.

The quality of the approximation is assessed for each cycle by independently using parameter $R^2$. The residuals graph may be also examined, see FIG.~(\ref{fig:residuals}). Usually cycles where pressure fluctuations are too large (i.e. $Max(PRS_B)-Min(PRS_B)>\SI{5e-3}{\milli\bar}$) are skipped because there is no way to find a good enough linear approximation. Most often these are the first 1-2 cycles after feeding water into the chamber. 

\includegraphics[width=\textwidth]{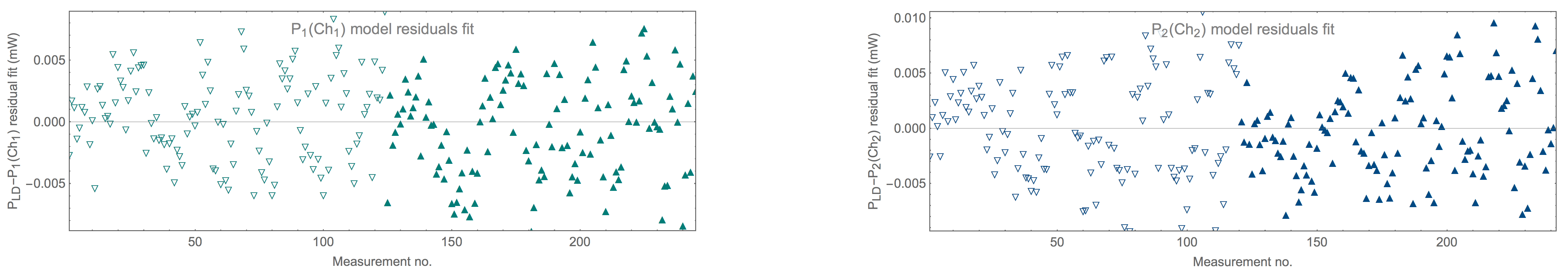}
\caption{Example of the linear regression residuals - only within the range of points designated for determining the regression.} \label{fig:residuals}
\end{figure*}

There are up to 960 (or 720) points ($Ch_{1n}$, $P_{LDn}$), ($Ch_{2n}$, $P_{LDn}$) available for each linear regression fitting. $P_{LD}$ is the power of light emitted by the laser as reported by the DFB built-in photodiode, $Ch_{1,2}$ is the detector's voltage reading, the $n$ index denotes the subsequent readings during scans. See FIG.~(\ref{fig:linear_approximation}).

The measurements obtained at both ends of the given cycle are used together to determine the detectors characteristics models. They always differ based on the direction of the laser current change (“up” or “down”). This way, any potential systematic shifts of the characteristics related to the laser temperature relaxation is eliminated. The triangles direction in FIG.~(\ref{fig:linear_approximation}) and FIG.~(\ref{fig:residuals}) denote measurements from two different directions of obtaining the characteristics.

We chose the parameters of the experiment (i.e. temperature and range of the laser current, absorption line, working pressure and length of the light path) in the way the characteristics are determined in a range where other water absorption lines are sufficiently weak. Therefore, water vapor do not cause any disturbances of the detector characteristics model calculations.

\subsection*{Transmittance Quotient calculation}
Gas optical transmittance is the ratio of the power of the light beam that has passed through the medium divided by the power of the light that has entered the medium. The characteristics of the detectors calculated above allow for calculating the transmittance for each of the channels:
\begin{equation}
\label{TR}
\begin{aligned}
 TR_1=P_1(Ch_1)/P_{LD}~, \\ TR_2=P_2(Ch_2)/P_{LD}~.
\end{aligned}
\end{equation}

Note that all “constant” light power loses (mainly resulting from mirrors reflectivity and photodiodes efficiency) are encoded within $P_{1,2}$ models. This way the Eq.~(\ref{TR}) are valid for determining just water vapor transmittance.

We compare transmittance values using the transmittance quotient ($TQ$) as defined below:

\begin{equation}
\label{TQ}
TQ=\frac{TR_1}{TR_2}= \frac{P_1(Ch_1)/P_{LD}}{P_2(Ch_2)/P_{LD}}=\frac{P_1(Ch_1)}{P_2(Ch_2)}~.
\end{equation}

\begin{figure*}[ht!]
\includegraphics[width=\textwidth]{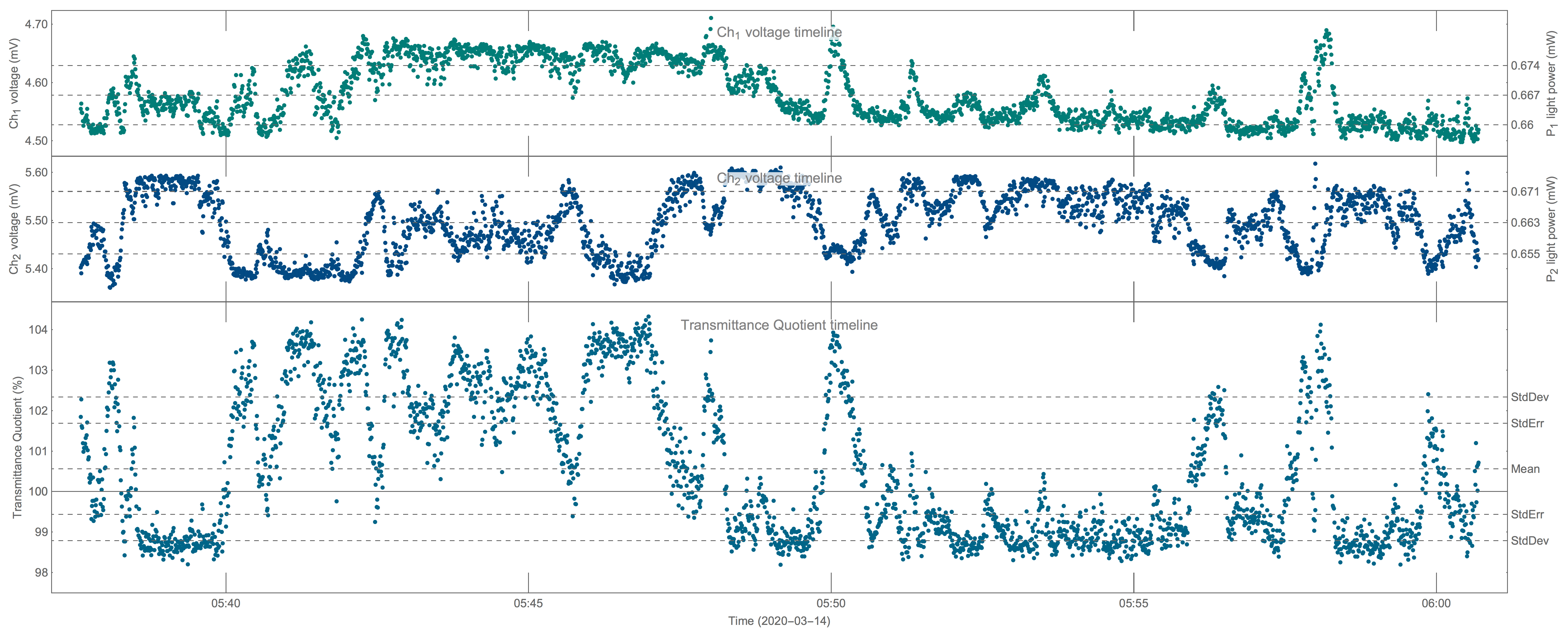}
\caption{Example of the $Ch_1$, $Ch_2$ and transmittance quotient ($TQ$) measurements within a single cycle.} \label{fig:tq}
\end{figure*}

There is either transmittance difference or transmittance quotient referred in this report. Transmittance difference is expressed as a percentage so it is just the transmittance quotient minus 1. $TQ$ values greater than 1 (transmittance difference greater than 0) mean that the transmittance measured by the smaller detector $Ch_1$ is greater than the transmittance measured by the classical one $Ch_2$. Examples of the $TQ$ values for the individual measurements are shown in FIG.~(\ref{fig:tq}). 

There is a possible issue with a non linearity of the DFB laser diode (current vs output power) and the DFB photodiode (incident light vs current). It was checked that this non linearity is so small in the operational range that it doesn’t significantly affect transmittance measurement. It should be taken into account for very accurate transmittance measurements. Besides we’re interested just in transmittance quotient in this experiment. This non linearity cancels out because it appears in both numerator and denominator of Eq.~(\ref{TQ}).

The actual reading (including its error) of the laser photodiode ($P_{LD}$) is irrelevant for the calculation of transmittance quotient $TQ$ because it cancels out appearing in both numerator and denominator of Eq.~(\ref{TQ}).

Determination of the transmittances and their quotient is possible only if the domains of characteristics $P_{1,2}$ includes the appropriate range of the $Ch_{1,2}$ voltages. It is checked automatically for each cycle. If the appropriate characteristic range is missing the measurements made within the given cycle are skipped by the aggregating algorithm. It may happen sometimes for sudden pressure changes.

\subsection*{Transmittance Quotient aggregation}

We determine the transmittance quotient for each single measurement within a cycle. There are 2 measurements per second recorded which makes hundreds measurements per cycle. Assuming that the experimental conditions are sufficiently constant during the cycle, the average transmittance quotient $\overline{TQ_k}$ for the entire cycle is determined:
\begin{equation}
\begin{aligned}
\overline{TQ_k}&=\frac{1}{M_k}\sum_{m=1}^{M_k}\frac{P_{1k}(Ch_{1m})}{P_{2k}(Ch_{2m})}\\
&=\frac{1}{M_k}\sum_{m=1}^{M_k}\frac{A_{1k}Ch_{1m}+B_{1k}}{A_{2k}Ch_{2m}+B_{2k}}\label{avgTQk}
\end{aligned}
\end{equation}

where $k$ - cycle index, $M_k$ – number of the single measurements in the $k$-th cycle, $m$ - measurement index within a cycle, $A_{1,2k}$, $B_{1,2k}$ – linear regression factors of ${P_{1,2k}}$ characteristics, respectively for channel 1 and 2.

Determination of the transmittance quotient $\overline{TQ}$ for the entire run (for single aperture) involves calculating the arithmetic mean of the quotients measured in the successive cycles:
\begin{equation}
\overline{TQ}=\frac{1}{K}\sum_{k=1}^{K}\overline{TQ_k}\label{avgTQ}
\end{equation}

where $K$ – number of cycles in a run.

\subsection*{Water vapor partial pressure calculation}

Having determined the classical transmittance in the $Ch_2$ channel, the partial pressure of water vapor ($PRS_{H_2O}$) can be determined based on the relationship between transmittance, absorbance and pressure:

\begin{equation}
PRS_{H_2O}=-0.00877\,Log(TR_2)\,\si{\milli\bar}~.\label{partialprs}
\end{equation}

We calculated the conversion factor depending on light path length, absorption line cross-section and temperature.

\section*{Results \& interpretation}

\begin{figure*}[!]
\includegraphics[width=\textwidth]{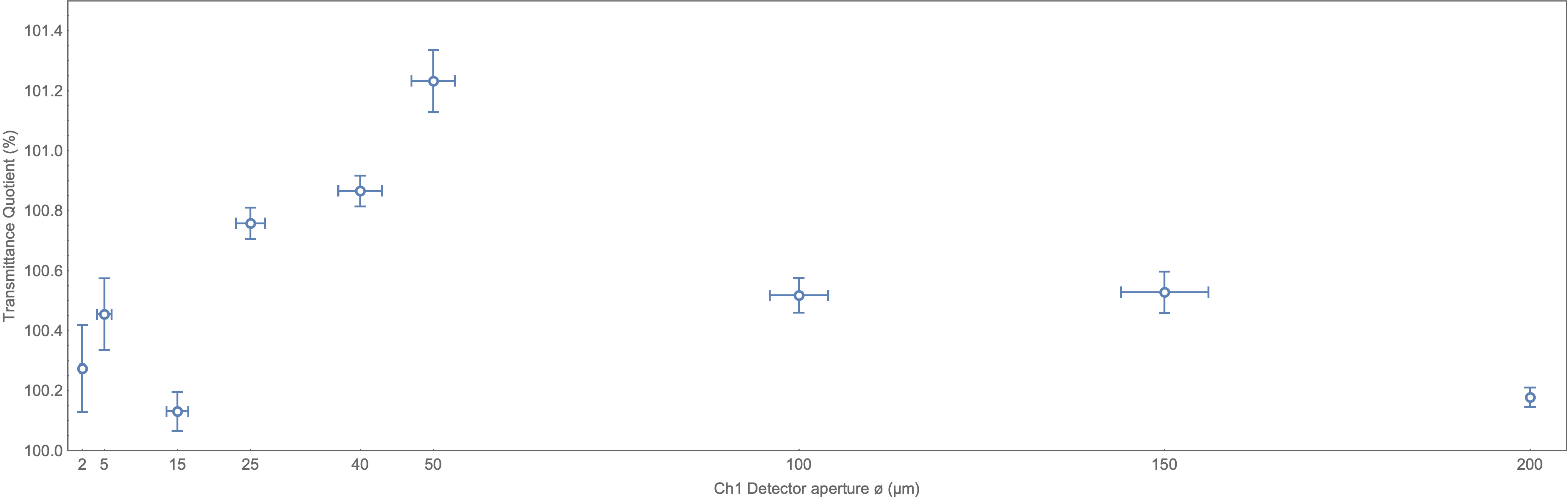}
\caption{The aggregated results of the experiment: mean transmittance quotients $\overline{TQ}$ along with their 1$\sigma$ standard errors (vertical bars) and aperture tolerances (horizontal bars). The transmittance quotients are averaged for all pressures for each examined aperture. The rightmost point is the result of no aperture runs.}
\label{fig:tq_size}
\end{figure*}

The aggregated results of the experiment are presented in FIG.~(\ref{fig:tq_size}). The mean transmittance quotient $\overline{TQ}$ is shown for each individual aperture examined. The rightmost point is the result of no aperture runs. All measured values are greater than 100\%. It means that the $Ch_1$ (smaller) detector measured higher transmittance.

The more detailed graphs showing the values of measurements for the individual cycles along with pressure and transmittance are presented in the Appendix, see FIG.~(\ref{fig:results}). Yet more data and figures are available online at www.smearedgas.org/experiment1. Among others, there are "cycle cards" prepared for all measurement cycles. They presents all essential data collected during single cycle along with parameters calculated, see Appendix for the guide. Raw data is available upon request. Most figures in this paper comes from the 30th cycle of run R-6, no. 20200305-0855 taken on Thu 5 Mar 2020 08:56.

\subsection*{Runs with aperture stops}

We observe that all runs with the pinhole installed show higher transmittance when measured by the smaller $Ch_1$ detector! This is exactly the  effect qualitatively predicted by the smeared gas theory. For runs with bigger, namely 25-, 40-, 50, 100- and  \SI{150}{\micro\metre} apertures, the statistical significance is higher than $5\sigma$. 

The maximum transmittance difference equals to 1.23\,$\pm$0.1\,\%, when the \SI{50}{\micro\metre} aperture is installed. For the 100- and \SI{150}{\micro\metre} apertures the difference is equal to 0.52\,$\pm$0.06\,\% and 0.53\,$\pm$0.07\,\% respectively. They are smaller than the \SI{50}{\micro\metre} aperture result. This is in line with the model's qualitative prediction: as the detector size increases, the measured transmittance should decrease - for detectors significantly larger then the wavelength (a kind of geometric optics approximation).

This relationship is the other way around for the 6 smallest apertures, however. Furthermore, for the 3 smallest apertures the statistical significance falls down to 2-3$\sigma$. This is because the transmittance, according to the smeared gas model, is proportional to the definite integral of $|\Psi|^2$ over so-called ”visibility tunnel”. The visibility tunnel is a volume where a photon amplitude doesn't cancel out when using certain detector - according to the path integrals approach \cite{Feynman2005}. For detectors comparable in size to the wavelength the non-cancelling photon probability amplitudes outside the "classic" visibility tunnel effectively thicken this tunnel. This increases the volume over which the probability distributions of the smeared gas particles are integrated. As a consequence, the likelihood of observing a photon scattering event with such a small detector rises. Therefore, the measured transmittance decreases - in the extreme case down to the classical transmittance level. We see this kind of relation on the left side of the FIG.~(\ref{fig:tq_size}). The transmittance quotient is decreasing towards 100\% along with decreasing $Ch_1$ detector aperture diameter (from 50- down to \SI{2}{\micro\metre}).

The result for the \SI{100}{\micro\metre} aperture (with much higher average pressure) is discussed later on. The very low reading for the \SI{15}{\micro\metre} aperture needs further analysis.

The quantitative model fitting to experimental results will be performed further down the line. Still, we observe qualitative predictions are met in the experiment.

\subsection*{Control runs without aperture}
During the R-8 and R-10 runs there are no apertures installed. Transmittance is measured using two naked detectors with a comparable diameter. The measured transmittance difference is 0.12\,$\pm$0.04\,\%. The transmittances are thus not perfectly identical but very close to each other. 

Apart from obvious measurement errors, such difference may result, for example, from the diameter tolerance due to the workmanship of the detectors themselves, impurities etc. This tolerance, in turn, is irrelevant in case of the measurements made with aperture stops in place because detector $Ch_1$ is obscured by the aperture anyway. Another reason for the observed difference may be a potential variance in the sensitivity of both detectors to different wavelengths as discussed earlier. This may cause some systematic error when determining the characteristics of all measurements. However, this control runs would indicate at least the order of magnitude of such a systematic error. Even if it is ${\sim}0.12$\,\%, it is still significantly smaller than the measured transmittance differences for 5 apertures larger than \SI{25}{\micro\metre}. Consequently, it does not undermine the conclusions drawn.

\subsection*{Transmittance quotient vs pressure}

There are at least two phenomena influencing the transmittance quotient that should be taken into account when the pressure changes and temperature is kept constant. As the pressure increases, i) the number of molecules increases (the transmittance drops), ii) the mean free path shortens. The experiment was not designed to examine these correlations so very little may be concluded in this field. However, there are some indications we can take a closer look at. They are in accordance with the model.

\subsubsection*{Transmittance quotient vs. number of molecules}
\begin{figure*}[!]
\includegraphics[width=\textwidth]{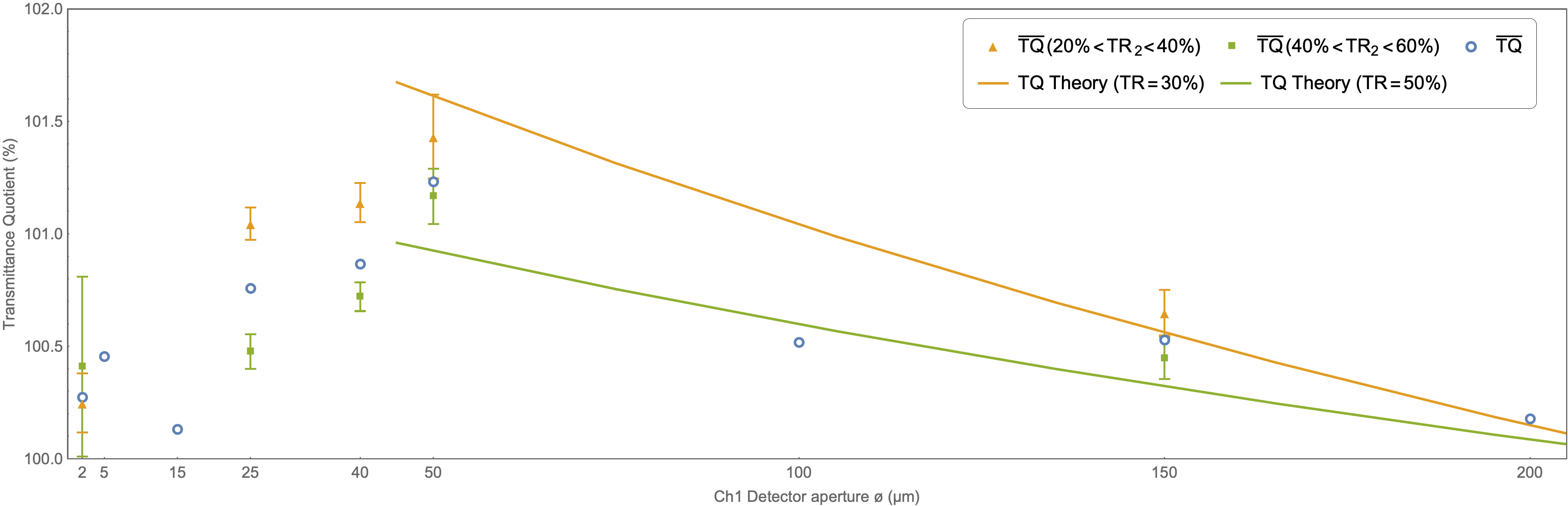}
\caption{A comparison of the experiment results with exemplary predictions of the model taking into account the basic pressure dependence. Solid lines represent exemplary predictions of the model in a range of "geometric optics" approximation (see text) for two different pressures (expressed as 30\% and 50\% transmittances). The measured transmittance quotients $\overline{TQ}$ averaged for two distinct $TR_2$ transmittances ranges (20-40\% and 40-60\% respectively) are presented with 1$\sigma$ standard error bars. The 5-, 15- and \SI{100}{\micro\metre} apertures are omitted due to missing data in this range of pressures. The transmittance quotients averaged for all pressures for all apertures are superimposed for convenience (without error bars).}
\label{fig:tq_press}
\end{figure*}

According to the model the increasing number of molecules with the mean wave function spread unchanged (aka the mean free time constant) along with a fixed visibility tunnel should increase the transmittance quotient. We show it qualitatively on the FIG.~(\ref{fig:tq_press}). There are presented sample predictions of the model in a range of "geometric optics" approximation for two different water vapor partial pressures: $\sim$\SI{1e-2}{\milli\bar} and $\sim$\SI{6e-3}{\milli\bar}. According to Eq.~(\ref{partialprs}) they correspond to transmittance ($TR_2$) of 30\% and 50\% respectively. The transmittances for each detector $d=1,2$ are calculated with the following equation \cite{Ratajczak2019b}:

\begin{equation}
TR_d(\bar{t}) = \prod_{n=1}^N \left( 1-\frac{G_d()}{4} \left[ erf \left( \frac{o_n-r_{Td}}{\sqrt2 stdev_{A_n}(\bar{t})} \right) - erf \left( \frac{o_n+r_{Td}}{\sqrt2 stdev_{A_n}(\bar{t})} \right) \right]^2 \right)~, \label{TRUpperLimit}
\end{equation}

where $\bar{t}$ denotes the molecules mean free time, $N$ is a number of molecules in the chamber, $o_n$ is the $n$-th molecule distance from the beam. The standard deviation $stdev_{A_n}(\bar{t})$ is assumed to be $\sim$\SI{14.6}{\micro\metre} (see next paragraphs). The geometry coefficient $G_d()$ is equal to cross section of the 101313–000212 [11] water absorption line ($\approx$\SI{7.76e-19}{\centi\metre^2}) divided by the area of the detector. Theoretical transmittance quotients are calculated assuming the transmittance tunnel has the shape of a \SI{60}{\metre} long truncated cone. The smaller base is of the size of an aperture. The angle (an effective divergence towards the source) is \SI{0.008}{\milli\radian}. Note, that this divergence is much smaller than the typical laser beam divergence thanks to the spherical mirrors re-focusing the beam for 80 times. The $r_{Td}$ is the radius of the (cylindrical) visibility tunnel, it's different for each detector $d$. For simplifying calculations the conical tunnel is approximated as a cylinder, thus $r_{Td}$ is equal to the mean of both bases radii.  

There are $TQ$ measurements shown on the FIG.~(\ref{fig:tq_press}) for at least 4 different apertures: 25-, 40-, 50- and \SI{150}{\micro\metre} following the model qualitative prediction: $\overline{TQ}$ is higher for the higher pressure. The statistical significance of this relation reaches as much as 3$\sigma$ for the 25- and \SI{40}{\micro\metre} apertures. Measurements for other apertures and pressures are omitted due to missing comparable data. 

\subsubsection*{Transmittance quotient vs. mean free path}
The measured transmittance quotient for the \SI{100}{\micro\metre} aperture stop is smaller than ones measured for 50- and \SI{150}{\micro\metre} apertures, see FIG.~(\ref{fig:tq_size}). According to the model $TQ$ closer to 100\% may indicate a shorter mean free time. Indeed for the \SI{100}{\micro\metre} aperture stop, most of the time the total pressure was higher than \SI{3e-2}{\milli\bar} and the water vapor partial pressure $PRS_{H_2O}$ was higher than \SI{1e-2}{\milli\bar}. The measurements for other apertures was made at a lower total pressure, approx. in a range from \SI{1e-3}{\milli\bar} to \SI{3e-2}{\milli\bar}. It seems that so high pressure eventually led to shortening the mean free path to a value smaller than the one constrained by the chamber size. This way the wave function standard deviation might drop below \SI{14.6}{\micro\metre} lowering the $\overline{TQ}$ reading.

\section*{Further works}
First of all it is necessary to conduct the quantitative model fitting. Careful consideration of the shape of the visibility tunnel is essential. All factors like reflections, collimation, laser power density or small apertures should be taken into account. 

The experiment should be repeated under modified conditions to eliminate any missed systematic errors and increase accuracy. Potential modifications include, among others:
\begin{itemize}
\item a split of the beam in parallel towards more than 2 detectors,
\item mounting apertures at each detector and testing the differences under varying configurations of reflection and beam split, e.g., 50/50 split,
\item better ways to reduce diode noise, e.g., changing the wiring or moving the first stage amplifiers closer to photodiodes, even inside the vacuum chamber,
\item more precise tuning of the wavelength to the absorption line using externally modulated laser driver, controlled with the standalone wavemeter or the lock-in feedback loop \cite{Neuhaus},
\item increasing the diameter of the vacuum chamber (extending the mean free path),
\item more stable fastening of the multi-pass cell mirrors,
\item removing the multi-pass cell rods shortening the mean free path near the laser beam,
\item applying a different number of reflections in the multi-pass cell,
\item assembling the measuring system outside the main section of the vacuum chamber so that the system could be reconfigured faster,
\item testing other gases e.g., methane (also with a large cross section in NIR).

\end{itemize}

A slight modification of the proposed method will allow for conducting measurements of transmittance differences and mean free paths for larger working pressure ranges. In particular, the length of the light path should be adjusted by controlling the number of reflections in the multi-pass cell. The experiment can be carried out without the use of a multi-pass cell, using a sufficiently long vacuum chamber.

Another modification may involve enabling limiting and adjusting the mean free path by installing internal walls or pipes. This will allow for studying the relationship between the transmittance quotient and the mean free path.

The presented values of the transmittance quotient in a range of 1\% may not seem respectable. However, higher values can be achieved e.g., by lengthen particles mean free time in a bigger chamber. 

\section*{Conclusions}
For ultra thin gas we observed the effect of changing the optical transmittance measurement readings using detectors of different sizes. For detectors of a diameter much bigger than the wavelength transmittance increased as detector size decreased. Thus, the objective of the experiment, which is to observe in laboratory conditions the qualitative effect predicted by the non-local smeared gas transmittance model, is achieved. The fact of observing a non-obvious phenomenon predicted by the theory is a strong premise for its validity. As far as we know there is no other gas transmittance model predicting such results.

The results are presented with a very high statistical significance level thanks to the long, repetitive measurement process. However, too few measurements were made to categorically determine the type of relationship between transmittance quotient and pressure. Moreover, there are many other improvements available regarding robustness, accuracy, flexibility etc. Some of them are outlined in the paper.

This experiment is a kind of a typical quantitative spectroscopy setup. It is relatively cheap and easy to repeat. However, it differs in two very important details from a well known TDLAS. Firstly, a sufficiently long gas particles mean free path is required. Free from any reactions that could be considered a quantum measurement or lead to decoherence. Therefore, the following should be provided: a large chamber, low pressure, electromagnetic shielding and weak measurement laser light. Secondly, the detector's light sensitive diameter should be comparable to the standard deviation of the smearing of the tested gas particles wave packets. It seems highly unlikely that both of these conditions could have been accidentally met during some earlier quantitative spectroscopy experiment. No reports of a similar experiment have been found in the literature.

In the experiment quite small detectors where used and a moderate pressure with a not so long light path was examined. However, the shown predictions of the smeared gas model apply for lower pressure levels, bigger detectors and longer distances. We believe the suitability of the smeared gas model should be verified in relation to the conditions of astronomical measurements.

\section*{Supporting information}

\begin{figure*}[!]
\includegraphics[width=\textwidth]{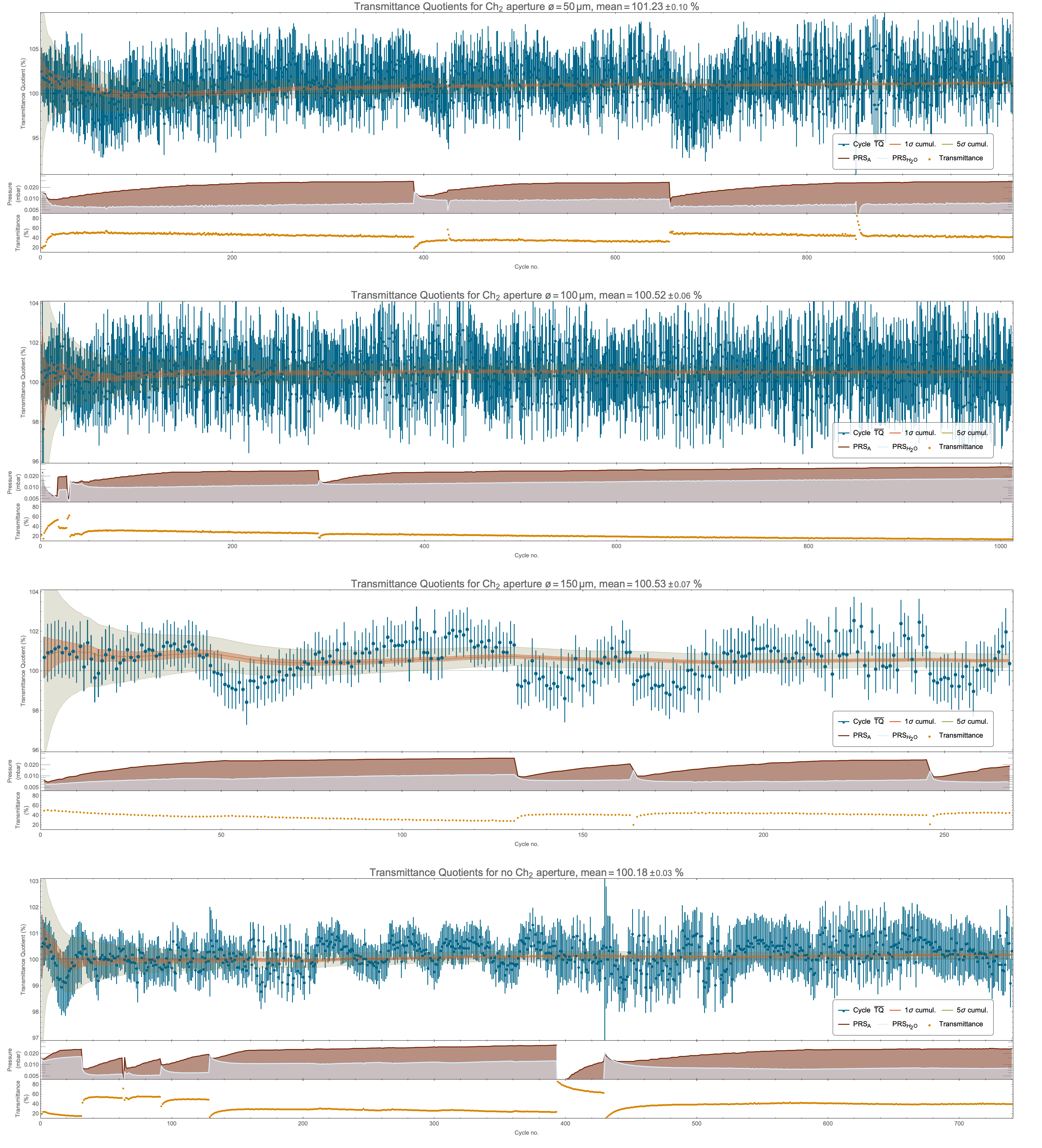}
\caption{Mean transmittance quotients along with their $1\sigma$ standard errors for the individual cycles and the bands showing the accumulated $1\sigma$ and $5\sigma$ measurement uncertainties, both total \& partial pressure and transmittance readings are superimposed.}
\label{fig:results}
\end{figure*}

\paragraph*{S1 Appendix.}
\label{S1_Appendix}
{\bf Cycle measurement results.}
There are measurement results for each cycle presented in FIG.~(\ref{fig:results}). They are grouped by the $Ch_2$ aperture size, aggregating data from different runs if required. Each point on a plot represents the measurement result for an entire (either 10- or 30-minutes long) cycle. The graphs also present bands showing the $1\sigma$ and $5\sigma$ measurement uncertainties accumulated for the preceding cycles. There are both total \& partial pressure and transmittance plots superimposed. Single measurements results are available online.

\begin{figure*}[!]
\includegraphics[width=\textwidth]{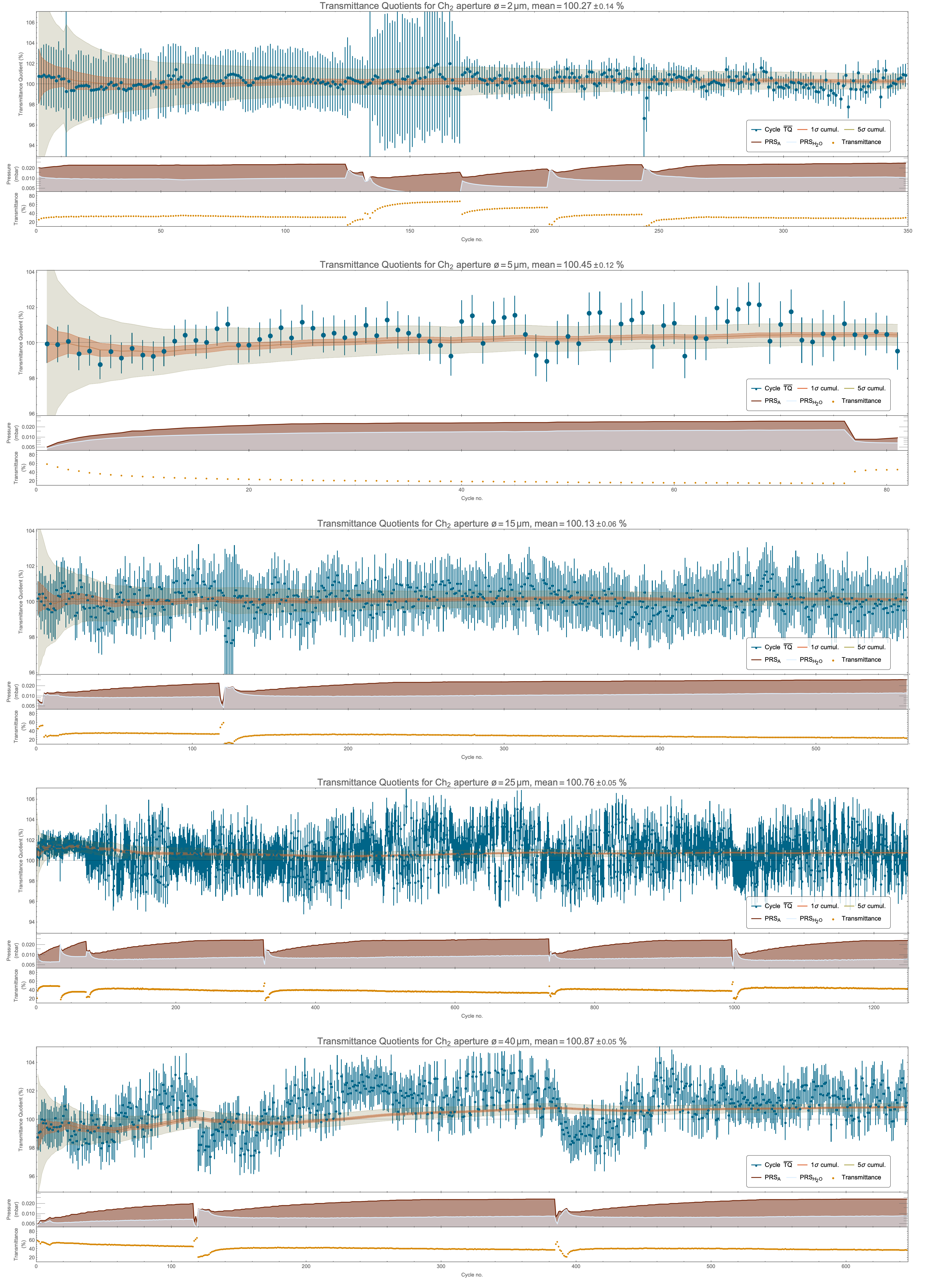}
\end{figure*}

\bigskip

\paragraph*{S2 Appendix.}
\label{S2_Appendix}
{\bf Online data.}

\begin{figure*}[]
\includegraphics[width=\textwidth]{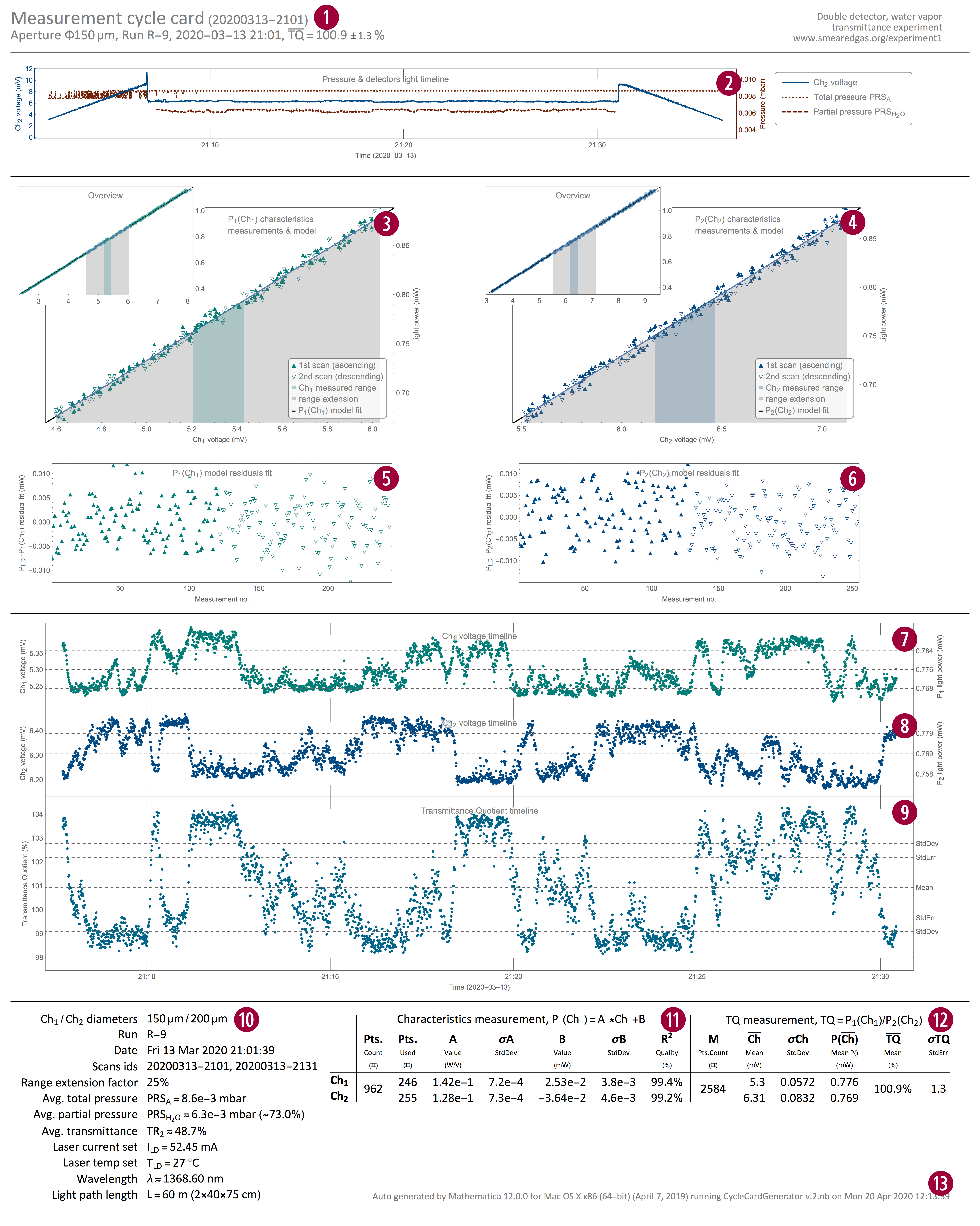}
\caption{All measurements taken during the cycle are shown on a sample cycle card. The numbers in the circles refer to the sections descriptions in the text.} \label{fig:cycle_card_sample}
\end{figure*}

Online data \& additional material are available on the website www.smearedgas.org/experiment1. There is a separate web page for each aperture size available. It consists of a run (or runs) summary, timeline figures and a list of measurement cycles within those runs. For each cycle on the list there is a link to a "cycle card". All single measurements taken during a cycle are shown on a single cycle card. A sample cycle card is presented in FIG.~(\ref{fig:cycle_card_sample}). It consists of the following sections.

\begin{enumerate}
\item The title of the card includes id of the cycle, the aperture diameter, the run name, timestamp the cycle started and the mean transmittance quotient ($TQ$) as measured in the cycle.
\item The timeline of the cycle contains the total \& partial pressure (red, dashed) and power of light (blue) as detected by the reference ($Ch_2$) detector. The partial pressure is calculated only for the period during which the transmittance measurement was carried out (the 3rd step of a cycle). The total pressure shown is measured with the $PRS_A$ vacuum meter.
\item The linear approximation of the $Ch_1$ characteristic along with source measurements. The colored vertical stripe indicates the range of the detected $Ch_1$ voltage variability during the cycle. The grey vertical stripes indicate an additional range (up to \SI{25}{\percent}) of the points used to determine the regression. The triangle direction reflects the laser current stepping direction during scanning (the 1st \& 4th steps of a cycle). Refer to FIG.~(\ref{fig:linear_approximation}) notes in the main text for more details. 
\item The same as the \#3 above but for the $Ch_2$ detector. 
\item The $Ch_1$ linear regression residuals - within the range of points designated for determining the regression.
\item The same as the \#5 above but for the $Ch_2$ detector. 
\item The timeline contains each single reading of the $Ch_1$ detector recorded. The left axis presents raw voltage as reported by the DAQ unit. On the right axis there is the corresponding power of light as calculated with the current cycle's $P_1$ characteristic model. The dashed horizontal lines denote the mean value and ${\pm}1\sigma$ deviation.
\item The same as the \#7 above but for the $Ch_2$ detector.
\item Each point is the transmittance quotient $TQ_m$ calculated for a pair of $Ch_{1m}$ \& $Ch_{2m}$ values from the same measurement. The dashed horizontal lines denote the mean value, its standard deviation and the standard error of the entire cycle.
\item A variety of the run and cycle parameters.
\item The parameters of both $P_1$ \& $P_2$ characteristics models determined in the cycle.
\item The mean transmittance quotient for the cycle (referred as $\overline{TQ_k}$ in the main text) along with its standard error and other uncertainty related parameters: the number of measurements, the mean detector readings and the relevant standard deviations.
\item The footer contains the software version used \& card generation timestamp.
\end{enumerate}

\paragraph*{S3 Appendix.}
\label{S3_Appendix}
{\bf Uncertainty and errors.}

Two main areas where errors originate can be distinguished in the system: i) errors in determining the characteristics of the detectors and ii) errors while performing the actual measurement of the transmittance i.e., power of light received by the detectors. The error in determining the characteristic is a systematic error in measuring the transmittance quotient in the given cycle. When aggregating data from multiple cycles, this error becomes a random error, however. Therefore, the rules used to calculate the standard error of the mean can be applied. Transmittance measurement errors are always random errors.

Although the setup remains unchanged all the time, the measurement error is different for the runs carried out with different apertures. This is due to the fact that for the individual apertures detector $Ch_1$ works at different parts of its characteristic. In particular, in case of small apertures of \SI{2}{\micro\metre} \& \SI{5}{\micro\metre} the detector works close to the lower limit of its sensitivity and therefore its signal to noise ratio is lower. In addition, various gains are used in channel $Ch_1$.

Furthermore, values of transmittance itself vary, as it depends, after all, on the partial pressure of the water vapor in the chamber. Error rate also depends on the wiring. The result is that the partial derivatives used to calculate error propagation in each cycle may vary. As a consequence, errors has to be estimated for each cycle independently.

The measure of uncertainty in the experiment is standard deviation and standard error. We use typical uncertainty propagation rules based on partial derivatives for their calculation \cite{Caria2001}.

Eq.~(\ref{avgTQk}) let us determine standard error $\sigma\overline{TQ_k}$ of the mean transmittance quotient in the $k$-th cycle: 

\begin{equation}
\left(\sigma\overline{TQ_k}\right)^2=\sum_{d=1}^{2}\left [
\left (\frac{{\sigma}A_{dk}}{\overline{P_{dk}}\,A_{dk}} \right )^2 +
\frac{1}{N_k}\left (\frac{{\sigma}Ch_{dk}}{\overline{P_{dk}}\,\overline{Ch_{dk}}} \right )^2 +
\left (\frac{{\sigma}B_{dk}}{\overline{P_{dk}}} \right )^2
\right ]~,\label{sigmaTQk}
\end{equation}

where aggregation based on $d=1,2$ denotes both detectors' errors. Parameters ${\sigma}A_{dk}$ and ${\sigma}B_{dk}$ denote standard deviations of parameters $A_{dk}$ and $B_{dk}$ of the characteristics models of detector $d$ in the $k$-th cycle. These deviations are calculated when linear regressions of characteristics $P_1(Ch_1)$ and $P_2(Ch_2)$ are determined.

There are some approximations applied in Eq.~(\ref{sigmaTQk}). They are based on the fact that throughout the entire cycle the $Ch_{dm}$ readings are relatively constant and approximately equal to the cycle's mean $\overline{Ch_{dk}}$:
\begin{equation}
\overline{Ch_{dk}}=\frac{1}{M_k}\sum_{m=1}^{M_k}Ch_{dm}~.
\end{equation}

Therefore $P_{dk}(Ch_{dm})$ can be approximated with $\overline{P_{dk}}$, an average independent of $n$:

\begin{equation}
\overline{P_{dk}}=P_{dk}\left(\overline{Ch_{dk}}\right)~.
\end{equation}

We also disregard the variability of ${\sigma}Ch_{dm}/Ch_{dm}$ for the successive measurements due to the low variability of the denominator. The ${\sigma}Ch_{dk}/\overline{Ch_{dk}}$ quotient is used instead, where ${\sigma}Ch_{dk}$ denotes the standard deviation of the measurements in detector $Ch_d$ during the $k$-th cycle. 

The $Ch_{dm}$ value may drift due to small pressure drift during the cycle. Usually it will be a small, continuous change due to a small continuous change of the partial pressure of water vapor. This drift has nothing to do with the measurement error and could be compensated (with a kind of trend detection) when determining the standard deviation inducted by real random errors. However, we don't compensate this trend what leads to a bit higher uncertainty. It's a conservative approach. It makes algorithm simpler and actually there is not so many cycles with high pressure drifts. It should be noted that the drift has no effect on the measurement of the transmittance quotient, since the quotient is calculated individually for each $m$-th measurement within a cycle. 

In order to determine standard error $\sigma\overline{TQ}$ for the entire measurement run we use the formula:
\begin{equation}
\sigma\overline{TQ}=\sqrt{\frac{1}{K}\sum_{k=1}^K\left(\sigma\overline{TQ_k}\right)^2}
\end{equation}

Preliminary analysis indicates following sources of noise during the experiment.
\begin{enumerate} 
\item Own noise of photodiodes operating at very low light with a significant dark current input.
\item Electromagnetic interference by photodiodes wiring.
\item Photodiodes and wiring noise amplification by both amplifier stages. 
\item Pendulum type vibrations of the upper mirror of the multi-pass cell amplified by the multiple reflections of the laser beam.
\item Laser light power and wavelength drift caused by laser current and operating temperature drifts.
\item Changes in polarization of light incident on the detectors resulting from optical fiber tension and vibrations.
\end{enumerate}

\section*{Acknowledgments}
The experiment was conducted at ACS Ltd. ACS Laboratory and the Institute of Plasma Physics and Laser Microfusion in Warsaw. We would like to thank Krzysztof Tomaszewski for enabling the experiment to be carried out at that site, for renting the vacuum equipment and supervision over vacuum evacuation as well as for his assistance. Special thanks go to Michał Popiel-Machnicki for documenting the project and for the operational support. We would also like to thank: Prof. Joanna Sułkowska, Prof. Tadeusz Stacewicz, Jarosław Baszak (Hamamatsu), Prof. Piotr Sułkowski, Prof. Janusz Będkowski, Prof. Piotr Wasylczyk, Prof. Dominik Jańczewski, Jacek Kaczmarczyk \& Grzegorz Reda for the subject matter discussions, providing the equipment \& resources as well as for their assistance.

The experiment was financed entirely from private funds and with the support of the above mentioned persons and organizations.

\nolinenumbers


\end{document}